\documentclass[sn-mathphys-num]{sn-jnl}

\usepackage{graphicx}%
\usepackage{multirow}%
\usepackage{amsmath,amssymb,amsfonts}%
\usepackage{amsthm}%
\usepackage{mathrsfs}%
\usepackage[title]{appendix}%
\usepackage{xcolor}%
\usepackage{textcomp}%
\usepackage{manyfoot}%
\usepackage{booktabs}%
\usepackage{algorithm}%
\usepackage{algorithmicx}%
\usepackage{algpseudocode}%
\usepackage{listings}%

\theoremstyle{thmstyleone}%
\theoremstyle{thmstyletwo}%

\theoremstyle{thmstylethree}%

\usepackage{amsmath}%
\begin{document}

\title[Article Title]{GRU-PFG: Extract Inter-Stock Correlation from Stock Factors with Graph Neural Network}


\author[1]{\fnm{Yonggai} \sur{Zhuang}}\email{start128@163.com}
\author[2]{\fnm{Haoran} \sur{Chen}}
\author[2]{\fnm{Kequan} \sur{Wang}}

\author*[1]{\fnm{Teng} \sur{Fei}}\email{feiteng@whu.edu.com}

\affil*[1]{\orgdiv{School of Resources and Environment Sciences}, \orgname{Wuhan University}, \orgaddress{\city{Wuhan}, \state{Hubei}, \country{China}}}
\affil*[2]{\orgdiv{Faculty of Science and Technology}, \orgname{BNU-HKBU United International College}, \orgaddress{\city{Zhuhai}, \state{Guangdong}, \country{China}}}




\abstract{The complexity of stocks and industries presents challenges for stock prediction. Currently, stock prediction models can be divided into two categories. One category, represented by GRU and ALSTM, relies solely on stock factors for prediction, with limited effectiveness. The other category, represented by HIST and TRA, incorporates not only stock factors but also industry information, industry financial reports, public sentiment, and other inputs for prediction. The second category of models can capture correlations between stocks by introducing additional information, but the extra data is difficult to standardize and generalize. Considering the current state and limitations of these two types of models, this paper proposes the GRU-PFG (Project Factors into Graph) model. This model only takes stock factors as input and extracts inter-stock correlations using graph neural networks. It achieves prediction results that not only outperform the others models relies solely on stock factors, but also achieve comparable performance to the second category models. The experimental results show that on the CSI300 dataset, the IC of GRU-PFG is 0.134, outperforming HIST's 0.131 and significantly surpassing GRU and Transformer, achieving results better than the second category models. Moreover as a model that relies solely on stock factors, it has greater potential for generalization.}




\maketitle

\section{Introduction}\label{sec1}

Stock investment is a significant means of wealth enhancement \cite{RN151}. By purchasing shares of relevant companies, investors become shareholders and derive income from the company's profits \cite{RN152}. In the long term, stock investment typically yields higher returns and serves as an effective hedge against inflation \cite{RN153}. However, stock investment also entails considerable risk. In scenarios of market volatility or poor corporate performance, investors face the possibility of losing their capital entirely \cite{RN154}. Risk mitigation and return enhancement are aspirations for many investors. Due to information asymmetry, the majority of investors often struggle to avoid risks, resulting in substantial losses \cite{RN155}.

Against this backdrop, to better mitigate risks and enhance returns, an increasing number of investors are focusing on the application of machine learning methods in stock prediction \cite{RN156, RN186}. Neural networks, exemplified by CNNs and Transformers, have opened new avenues for stock forecasting \cite{RN158}. Although the predictions inevitably deviate to some extent from reality, a well-designed predictive model can significantly aid investors in formulating their stock selection strategies \cite{RN159}.

Current stock prediction models can be categorized into two types. The first type relies solely on stock factors as inputs for prediction, without incorporating additional information. The second type considers not only stock factors but also integrates extra information such as stock relationships, market capitalization, and industry financial reports.The former type, represented by models such as GRU (Gated Recurrent Units) and GATs (Graph Attention Networks), uses relatively objective and standardized information \cite{RN160}. However, previous research often suggests that these methods lack sufficient information, making it challenging to achieve more efficient predictions. The latter type, exemplified by models such as MDGNN and HIST, introduces information like industry relationships and stock relationships. These methods place each stock within the industry framework, calculating the correlations between different stocks to predict returns more effectively \cite{RN162, RN161}.

Previous research generally suggests that the latter type of model, which utilizes more comprehensive information, achieves higher accuracy \cite{RN163}. However, this approach also has significant drawbacks. Firstly, information such as industry concepts, market capitalization, and financial reports is abundant (difficult to select) and often subjective, lacking objective standards for determining which data should be used as model inputs \cite{RN164}. Secondly, this information tends to lag, meaning investors often realize critical industry insights only after significant stock changes have already occurred \cite{RN165}. In other words, there are frequent instances where the stock has undergone substantial movements before investors recognize the relevant industry information. Additionally, the strategies of many companies are rapidly evolving, and new companies are constantly emerging \cite{RN166}. For these new strategies and companies, the method based on historical industry relationships faces the risk of becoming obsolete. This means that predictions relying on past industry relationships may not be effective in these rapidly changing environments.

Considering the advantages and disadvantages of the two types of models, although the latter has shown good accuracy in some research, the issues it has suggest that the former has broader application prospects. Based on this, the present study references related models that rely solely on stock factors. The MLP (Multilayer Perceptron) is the most basic model, learning factor information through multiple dimensionality reductions and nonlinear processing \cite{RN167}. However, its ability to learn relationships between factors is limited due to the constrained capacity of its network. GRU (Gated Recurrent Unit), LSTM (Long Short-Term Memory), ALSTM (Attention-based LSTM), and Transformer models treat the relationships between factors as temporal sequences, enabling better extraction of factor information, representing a significant improvement over MLP \cite{RN168, RN169}. Nonetheless, these models do not perform well in capturing relationships between different stocks, which limits their prediction performance. 

While numerous models rely solely on stock factors, few focus on the roles of stocks within the market, and even fewer consider inter-stock relationships. \cite{RN171}. Meanwhile, the correlations between stocks contain a wealth of information that can significantly help to stock prediction. Based on this, we propose a new stock prediction model purely based on stock factors, named GRU-PFG (Project Factors into Graph). In our approach, we project the factor information processed by the GRU onto a Graph Network to learn the relationships between stocks. With the Graph Network, we integrate factor information with inter-stock relationships and achieve better prediction performance.  In summary, this study makes the following major contributions:

\begin{enumerate}[1.]
\item We have innovatively proposed a more effective stock prediction model, GRU-PFG. Among prediction models purely based on Alpha360 stock factors, GRU-PFG outperforms previous models.
\item GRU-PFG deeply extracts the information embedded in Alpha360. Compared to models like ALSTM and GRU, GRU-PFG provides more profound extraction. This model offers a new approach for further mining stock factor information.
\item Compared to prediction models based on multiple sources of information (such as HIST and TRA), our model achieved results that are on par with these models in experiments. In this context, our model provides an approach with greater potential for generalization, as it can extract inter-stock correlations from stock factors.
\end{enumerate}

In summary, the structure of this paper is as follows: In the second chapter, we discuss related works in stock prediction, highlighting the importance of the prediction model based on pure Alpha360 factors. In the third chapter, we introduce some foundational work of our model, including basic concepts related to stocks and the application of GRU networks in stock prediction. In the fourth section, we present the basic framework and algorithms of our model, and in the fifth section, we validate the superiority of our model through experiments on various metrics such as IC (Information Correction), Rank IC, and precision@N. Finally, we summarize and provide an outlook on our work.

\section{Related Work}\label{sec2}
\subsection{Stock Factors}\label{subsec1}

The quantitative work related to stock prediction has long been shrouded in secrecy due to the difficulty of obtaining insider information related to trading \cite{RN172}. Against this backdrop, there has been a long-standing desire to establish a stable and reliable model for quantitative stock work using publicly available data such as price and volume \cite{RN173}. Some quantitative analysts set up a series of patterns based on stock trends and use their experience to judge the next move of the stock \cite{RN174}; while others have further quantified and standardized the data, proposing stock factors for stock prediction. 

In 2015, Zurab Kakushadze and Igor Tulchinsky \cite{RN175} defined stocks with 4000 numerical values and explored the correlation between future stock changes and these 4000 numerical values, marking the first step in the application of stock factors. However, the 4000 numerical values were too redundant, so Zura Kakushadze \cite{RN176} later proposed using 101 numerical values to quantify stocks and verified the relationship between these 101 factors and stock fluctuations. Xiao Yang, et al. \cite{RN177}, considering the role of stock factors in stock information, referenced previous research and introduced stock factors into the qlib platform, setting up two types of stock factors: Alpha360 and Alpha158. As the qlib platform has gained increasing attention, Alpha360 and Alpha158 factors have been widely adopted by models such as GRU, GATs, Transformer, HIST, and TRA, becoming the standard input for machine learning (deep learning) methods to predict daily stock trends \cite{RN179, RN178, RN161}.

\subsection{Prediction Models Purely Based on Alpha360}\label{subsec2}

Among the two types of factors provided by the qlib platform, Alpha360 stock factors have more dimensions and are more suitable for processing with deep learning methods such as GRU, LSTM, and Transformer models \cite{RN180}. In the application of deep learning, most models process based solely on stock factors. Compared to methods that based on multiple information, this approach is more standardized and universal.

For Alpha360 factors, MLP is the most basic model. It learns the factors through multiple dimensionality reductions and nonlinear processing, but its ability to learn the relationships between factors is limited by its network capacity \cite{RN181}. GRU \cite{RN182}, LSTM \cite{RN183}, ALSTM \cite{RN184}, and Transformer \cite{RN185} models treat the factor information as temporal sequences, allowing for better extraction of factor information, achieving a certain breakthrough compared to MLP. However, these models do not perform well in capturing the relationships between different stocks, which limits their performance \cite{RN161}.

\subsection{Prediction Models Based on Multiple Sources of Information}\label{subsec3}

In recent years, some scholars have argued that the information contained in stock factors is limited and have begun to consider incorporating additional information into the stock prediction process \cite{RN186}. This means use not only stock factors (trends of stocks over a certain period) as model inputs but also other information as predictive evidence.

Some scholars have considered public sentiment, believing that sentiment information related to stocks might contain insights into how industries are affected, thereby predicting future trends \cite{RN187}. Other scholars have suggested that information such as company market capitalization and industry concepts may imply inter-stock relationships, and representing these relationships through graphs can further improve prediction accuracy \cite{RN161, RN188}. Additionally, some scholars have delved into company-provided data such as industry reports and earnings reports to uncover intrinsic information about individual stocks, supplementing stock factor information \cite{RN189}.

Prediction models based on multiple sources of information can potentially enhance prediction performance to some extent. However, they also face several problems:

\begin{enumerate}
\item The selection of multiple sources of information is subjective and difficult to standardize and unify.
\item Even with multiple sources of information, there is still the issue of information lag. Multiple sources of information cannot guarantee better performance across different datasets at different times.
\item There is a lack of information for new companies and emerging industries.
\end{enumerate}

\begin{table}[h]
\caption{Characteristics of two different stock prediction models}\label{tab1}%
\begin{tabular}{@{}p{3cm} p{3cm} p{8cm}@{}}
\toprule
Method Type & Representative Models  & Feature \\
\midrule
Model only based on Alpha360 factors & ALSTM, GRU, GATs and GRU-PFG (ours)   & More standardized and universal but need to find a good way to extract more information. \\
Predict by multiple sources of information  & TRA, HIST and MDGNN & Bringing in more information may help to predictions, but the selection of multiple information sources is subjective and challenging to standardize or unify. \\

\botrule
\end{tabular}
\end{table}

\section{Fundmentals}\label{sec3}

\subsection{Stock trend-related concepts}\label{subsec3}

Stock prices are influenced by various factors, including a company's financial performance \cite{RN190}, market sentiment \cite{RN191}, and macroeconomic conditions \cite{RN192}. Although stock prices result from multiple influencing factors, they tend to change smoothly and continuously within a certain time frame in the stock market, rather than experiencing sudden jumps or interruptions. By analyzing the changes in stock prices over a past period, one can objectively reflect the combined effects of these influencing factors to some extent and further predict the stock price trend for the next stage \cite{RN193}.

Generally, stock trends are primarily analyzed by comparing and calculating changes in stock price data. If the stock price at time t is  
${P_t}$ and the stock price at time t+1 is ${P_{t + 1}}$, then the change rate over this period can be used to define the stock's trend as follows:

\begin{equation}
\alpha  = \frac{{{P_{t + 1}} - {P_t}}}{{{P_t}}}
\end{equation}

From the stock's variation trend, we can further derive the daily stock return. If the opening price of a stock on a given day is ${P_{open}}$ and the closing price is ${P_{close}}$, then the stock's return for that day is given by:

\begin{equation}
\beta  = \frac{{{P_{close}} - {P_{open}}}}{{{P_{open}}}}
\end{equation}

In the field of financial investment, Alpha360 factor is commonly used to define the changes in stock prices over a period (60 days) to predict the trend of stock price movements in the next phase. Alpha360 factor is an objective metric composed of various pieces of information from the stock market, used to quantify the difference between the expected return of a stock and the overall market return. It represents a collection of variables that may affect stock returns, including fundamental factors, technical indicators, market sentiment, and macroeconomic data.
The concept of the Alpha360 factor embodies the idea of multi-factor methods, which consider not just a single factor but a combination of multiple factors when predicting future stock or portfolio returns. This approach aims to capture various dimensions of stock returns to improve prediction performance.

\subsection{Limitations of GRU in Stock Prediction}\label{subsec6}

In general, neural networks handling daily stock data typically use 360 factors as inputs to the prediction network, with the output being the return of one day. While pure GRU networks may not extract information from the 360 factors as effectively, they still demonstrate relatively good performance compared to networks like Transformer and LSTM in daily predictions \cite{RN161}.

\begin{figure}[h]
\centering
\includegraphics[width=0.9\textwidth]{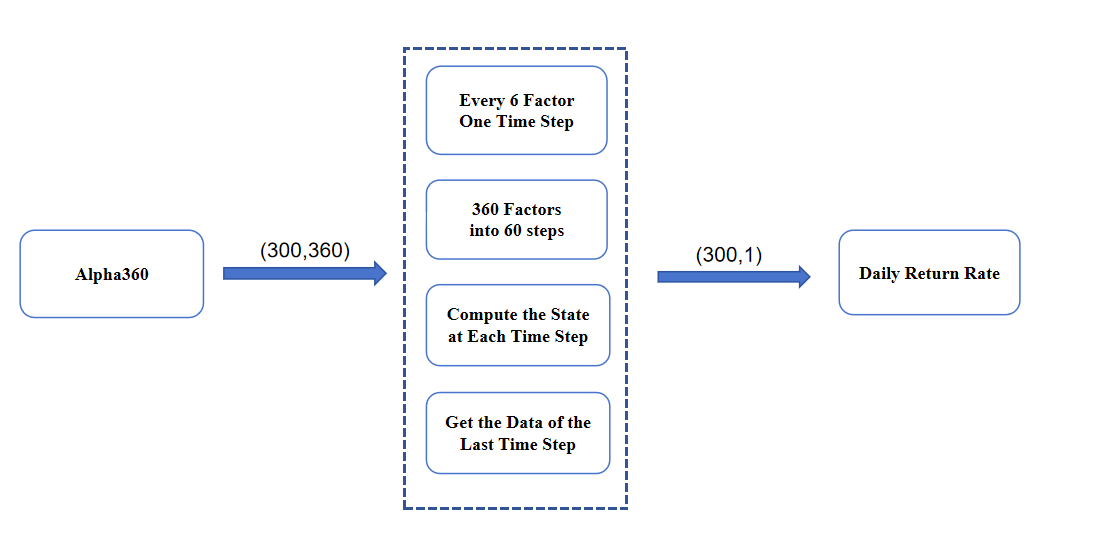}
\caption{GRU processing workflow on CSI300 data: Input 360 factors, output one-dimensional daily return}\label{fig1}
\end{figure}

In the prediction process with GRU networks, time sequence information is required. During processing, every 6 factors out of the 360 factors are treated as a single time step (60 steps in total). After processing, the data from the final time step is used to reduce the dimensionality of the 360-dimensional data. Finally, a Linear function is applied to predict the 1-dimensional return data. This prediction process allows GRU to extract information through stepwise dimensionality reduction, making it straightforward and efficient.

However, in the stock market, the hidden relationships between stocks and the roles of stocks in the market are significant factors affecting stock price movements. GRU's ability to extract and utilize inter-stock relationship information is poor, which limits its effectiveness in capturing these crucial factors.

\section{Methodology}\label{sec4}

Considering the limitations of previous models, we propose a new model, GRU-PFG (only based on Alpha360), to achieve improved stock prediction performance.

Our proposed stock prediction network, GRU-PFG, is divided into three parts: (1) Preliminary Information Extraction (2) Primary Relationship Extraction (3) Secondary Relationship Extraction. We use the Alpha360 factors of multiple stocks as input to predict the returns of these stocks.

\begin{figure}[h]
\centering
\includegraphics[width=0.9\textwidth]{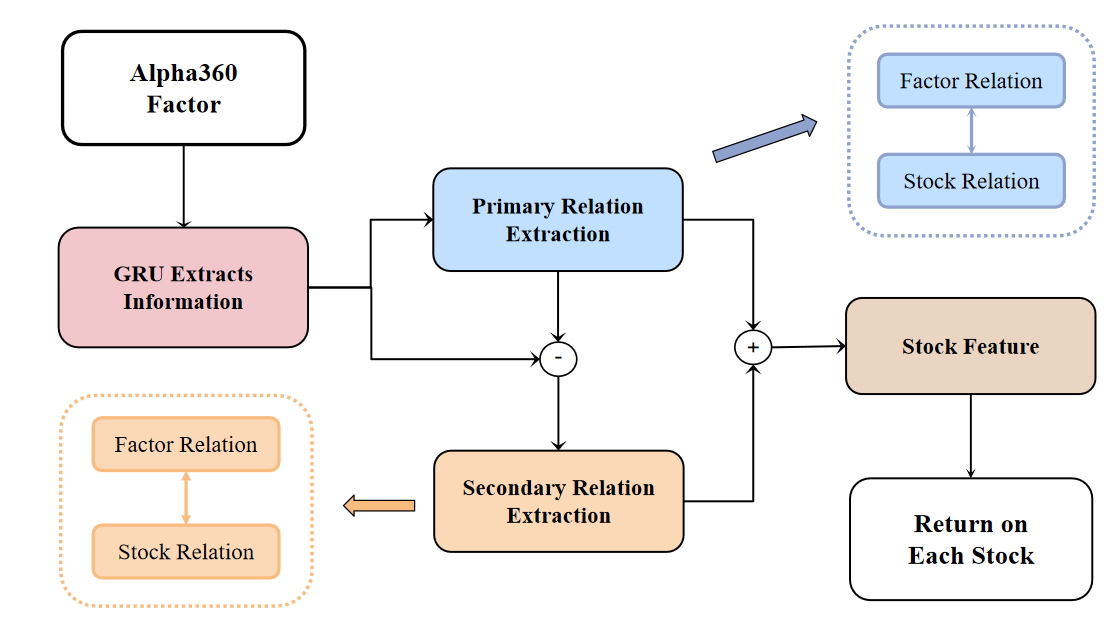}
\caption{Prediction Workflow of GRU-PFG model}\label{fig2}
\end{figure}
\noindent\textbf{Preliminary Information Extraction.} In the preliminary information extraction phase, we use a GRU network. With 360 factors as input, we treat every 6 factors as one time step, reducing the 360-dimensional information to a 64-dimensional representation for subsequent processing. During this stage, the relationships between different factors  are extracted. After GRU processing, Stock factor information is extracted. and feature representation of each stock is obtained:

\begin{equation}
X = \left( {\begin{array}{*{20}{c}}
{{X_{{\rm{a1}}}}}&{{X_{{\rm{a2}}}}}& \cdots &{{X_{{\rm{an}}}}}\\
{{X_{{\rm{b1}}}}}&{{X_{{\rm{b2}}}}}& \cdots &{{X_{{\rm{bn}}}}}\\
 \cdots & \cdots & \cdots & \cdots \\
{{X_{m1}}}&{{X_{m2}}}& \cdots &{{X_{{\rm{mn}}}}}
\end{array}} \right)
\end{equation}
where ${X_{{\rm{mn}}}}$ represents the information for the m-th stock and the n-th dimension after processing.
\\\\
\noindent\textbf{Primary Relationship Extraction.} After obtaining the feature representation information for each stock from the GRU processing, we apply the softmax function to X along the horizontal and vertical dimensions to obtain matrices ${X_ - }$ and ${X_|}$, respectively. These matrices represent the roles of each value ${X_{{\rm{mn}}}}$ in the m-th stock's internal dimension and the n-th dimension across different stocks.

The matrices ${X_ - }$ and ${X_|}$, maintain the same form as X, where  ${X_ - }$ represents the relationships between internal factor information within each stock, and ${X_|}$ represents the correlations between different stocks. We then compute the Pearson correlation coefficient for  ${X_ - }$ and ${X_|}$ across the dimensions of each stock to obtain the matrix R: 

\begin{equation}
    {R_{{\rm{xy}}}} = \frac{{\sum _{{\rm{i}} = 1}^{\rm{n}}({F_{xi}} - \overline {{F_x}} )({F_{yi}} - \overline {{F_y}} )}}{{\sqrt {\sum _{{\rm{i}} = 1}^{\rm{n}}{{({F_{xi}} - \overline {{F_x}} )}^2}} \sqrt {\sum _{{\rm{i}} = 1}^{\rm{n}}{{({F_{yi}} - \overline {{F_y}} )}^2}} }}
\label{equ4}
\end{equation}
where x represent the x-th stock in matrix ${X_ - }$, y represent the y-th stock in matrix ${X_|}$, i represent the i-th factor information for each stock, $\overline {{F_x}} $ represent the mean factor information of the x-th stock in matrix , and $\overline {{F_y}} $ represent the mean factor information of the y-th stock in matrix.

We compute the aggregated weight ${R_{{\rm{1}}}}$ by using the Pearson correlation coefficients. After that, we multiply ${R_{{\rm{1}}}}$ with X, the to obtain the Stock Feature derived from the primary relationship extraction. In this process, our approach is significantly more complex than traditional GRU networks. This complexity is necessary to calculate the relationships between each factor and other factors, as well as between each stock and other stocks. To achieve this, we  construct the matrix ${R_1}$ as a correlation relationship matrix to serve as the feature weight matrix.
\\\\
\noindent\textbf{Secondary Relationship Extraction.} In financial markets, there are complex industry and trading relationships. We believe that beyond the primary relationships, there are additional latent relationships. These latent relationships, together with the primary relationships, form the following expression:

\begin{equation}
    {X_{hid}} = X - {W_a}{X_ - } - {W_b}{X_|}
\end{equation}
where X is the initial information extracted by the GRU, ${X_ - }$ and ${X_|}$ represent the information extracted in the primary relationship phase, ${W_a }$ and ${W_b }$ are learnable parameters, ${X_{hid}}$ is the information to be extracted in the secondary relationship phase. We apply similar operations as in the primary relationship extraction: using softmax and Pearson correlation coefficients, we construct an aggregation matrix  ${R_2}$ and multiply it with ${X_{hid}}$ to obtain the Stock Feature in the secondary relationship phase.
\\\\
\noindent\textbf{Stock Feature.} The final Stock Feature  is ${F_{last}}$ obtained by combining the Stock Features derived from the above two relationship extraction phases:

\begin{equation}
    {F_{last}} = {W_c}X + {W_d}{R_1}X + {W_e}{X_{hid}} + {W_f}{R_2}{X_{hid}}
\end{equation}
where ${W_c}$, ${W_d}$, ${W_e}$ and ${W_f}$ are learnable parameters, ${F_{last}}$ is the final feature used to predict the change trends of all stocks:

\begin{equation}
    {p^t} = {W_l}{F_{last}} + {b_l}
\end{equation}
where ${W_l}$ and ${b_l}$ are learnable parameters, 
${p^t}$ represents the predicted daily return on date t.
\\\\
\noindent\textbf{Loss Function.} The loss function L for GRU-PFG is implemented using Mean Squared Error (MSE) and has the following form:

\begin{equation}
    L = \sum\limits_{t \in D} {\sum\limits_{i \in {S^t}} {\frac{{{{(p_i^t - g_i^t)}^2}}}{{\left| {{S^t}} \right|}}} }
\end{equation}
where D represents the set of dates, t represents a specific date, ${S^t}$ represents the set of stocks on date t, ${p_i^t}$ represents the predicted daily return for the i-th stock on date t, ${g_i^t}$ represents the actual daily return for the i-th stock on date t.

\section{Experiments}
\subsection{Experiment Setup}

We test our model based on the CSI100 and CSI300 datasets. We use stock data from January 1, 2007, to December 31, 2014, as the training set, stock data from January 1, 2015, to December 31, 2016, as the validation set, and stock data from January 1, 2017, to December 31, 2020, as the test set. We calculate the Alpha360 for different times as model inputs using the qlib platform, and use the predicted daily return rates as model outputs with the actual daily return rates as labels. To highlight the performance of our model, we compare the GRU-PFG model with other models  under the same dataset, comparing metrics such as IC (The Information Coefficient), rank IC, and Precision@N in the test set.

The Information Coefficient is an important metric for evaluating stock prediction models. During training and testing, we calculate the Pearson correlation coefficient between the model’s predictions and the actual labels. During training, IC is used as the loss function, while during testing, a higher coefficient indicates better prediction performance. Rank IC, or Spearman’s rank correlation coefficient, measures the correlation between the predicted ranks and actual ranks of stocks, with a higher coefficient indicating better prediction accuracy. Additionally, following the work of other researchers, we also consider Precision@N as a key evaluation metric. We calculate the proportion of stocks in the top N predicted stocks that are actually in an upward trend. A higher Precision@N indicates that our model better aids investors in improving returns.

In the experiment, our proposed GRU-PFG model is compared against two types of models. The first category includes MLP, ALSTM, GRU, SFM, GATs, and Transformer, which solely rely on the stock factor Alpha360 for prediction. The second category comprises TRA and HIST, which, in addition to stock factor information, incorporate industry financial reports, market trading data, and other inputs. All comparison models are implemented on the qlib platform.

In previous studies, HIST has been recognized as the most effective prediction model. However, it requires multi-source input data, making standardization and generalization challenging. Among models purely based on the Alpha360 stock factor, GRU and ALSTM are considered the most effective.

\begin{table}[h]
\centering
\caption{Experimental setup for the GRU-PFG model in stock prediction}
\begin{tabular}{|l|l|}
\hline
\textbf{Dataset} & CSI100 and CSI300 \\ \hline
\textbf{Comparison Model} & 
MLP, ALSTM, GRU, SFM, GATs and Transformer (just based on Alpha360), \\ & TRA and HIST (multi-source) \\ \hline
\textbf{Evaluating Indicator} & IC, Rank IC and Precision@N \\ \hline
\textbf{Model Input} & Alpha360 \\ \hline
\textbf{Model Output} & Daily return of each stock \\ \hline
\end{tabular}
\end{table}

\subsection{Evaluation Results}

The experimental results based on the CSI100 and CSI300 datasets demonstrate a significant advantage of our model over previous models (which only use Alpha360 factors for prediction). GRU-PFG performs exceptionally well across various metrics such as IC, Rank IC, and Precision@N (See Table \ref{table3}). 

Notably, Our model outperforms the previous sota model HIST on the CSI300 dataset. It's important to note that our model uses only stock factors as input, while HIST incorporates additional industry information and corporate valuations beyond stock factors. Our model achieves better prediction results with fewer inputs. This performance also indirectly highlights the effectiveness of the network in extracting inter-stock correlations.

It is worth mentioning that our model exhibits a greater advantage on the CSI300 dataset compared to the CSI100 dataset. One possible explanation is that our model effectively captures the relationships between stocks. As the number of stocks increases, our model can capture more related information, naturally leading to improved performance.

\begin{sidewaystable}[htbp]
\centering
\small 
\caption{Comparison of Methods on CSI 100 and CSI 300}
\begin{tabular}{cccccccccccccc}
\hline
\multirow{2}{*}{\shortstack{Methods only based \\ on Alpha360}} & \multicolumn{3}{c}{CSI 100} & \multicolumn{6}{c}{CSI 300} \\
 & IC (↑) & Rank IC (↑) & \multicolumn{3}{c}{Precision@N (↑)} & & IC (↑) & Rank IC (↑) & \multicolumn{3}{c}{Precision@N (↑)} \\
 & & & 3 & 5 & 10 & 30 & & &  3 & 5 & 10 & 30 \\
\hline
MLP & 0.071 & 0.067 & 56.53 & 56.37 & 55.49 & 53.55 & 0.082 & 0.079 & 57.21 & 57.15 & 55.55 & 55.36 \\
(just Alpha360) & (4.8e-3) & (5.2e-3) & (0.51) & (0.48) & (0.30) & (0.08) & (6e-4) & (3e-4) & (0.39) & (0.33) & (0.34) & (0.14) \\
LSTM & 0.097 & 0.091 & 60.12 & 59.49 & 59.04 & 54.77 & 0.104 & 0.098 & 59.27 & 58.40 & 56.98 & 55.01 \\
(just Alpha360) & (2.2e-3) & (2.3e-3) & (0.32) & (0.31) & (0.43) & (0.14) & (1.5e-3) & (1.4e-3) & (0.42) & (0.38) & (0.40) & (0.12) \\
GRU & 0.103 & 0.097 & 59.96 & 58.37 & 55.95 & 54.55 & 0.113 & 0.108 & 59.28 & 58.57 & 57.43 & 55.37 \\
(just Alpha360) & (1.7e-3) & (1.8e-3) & (0.49) & (0.33) & (0.45) & (0.25) & (1.1e-3) & (1.2e-3) & (0.43) & (0.51) & (0.35) & (0.29) \\
SFM & 0.081 & 0.074 & 57.79 & 56.52 & 53.95 & 53.88 & 0.102 & 0.096 & 58.84 & 58.07 & 56.82 & 55.28 \\
(just Alpha360) & (7.0e-3) & (8.1e-3) & (0.65) & (0.44) & (0.42) & (0.21) & (3.2e-3) & (2.7e-3) & (0.46) & (0.38) & (0.30) & (0.15) \\
GATs & 0.096 & 0.090 & 59.17 & 58.71 & 58.13 & 54.79 & 0.111 & 0.105 & 60.49 & 59.96 & 57.41 & 56.39 \\
(just Alpha360) & (4.5e-3) & (4.4e-3) & (0.68) & (0.52) & (0.47) & (0.39) & (1.9e-3) & (1.9e-3) & (0.50) & (0.49) & (0.29) & (0.21) \\
ALSTM & 0.102 & 0.097 & 59.41 & 58.33 & 55.32 & 55.02 & 0.115 & 0.109 & 59.31 & 58.52 & 57.47 & 57.10 \\
(just Alpha360) & (1.8e-3) & (1.9e-3) & (0.39) & (0.34) & (0.40) & (0.09) & (1.1e-3) & (1.2e-3) & (0.52) & (0.44) & (0.37) & (0.20) \\
Transformer & 0.089 & 0.090 & 59.13 & 58.48 & 55.98 & 55.01 & 0.106 & 0.103 & 60.06 & 59.30 & 57.11 & 56.14 \\
(just Alpha360) & (6.4e-3) & (6.0e-3) & (0.36) & (0.39) & (0.43) & (0.11) & (3.3e-3) & (3.0e-3) & (0.46) & (0.39) & (0.35) & (0.27) \\
HIST & 0.120 & 0.115 & 61.87 & 60.82 & 59.38 & 56.04 & 0.131 & 0.126 & 61.60 & 61.08 & 60.51 & 58.79 \\
(multi-source) & (1.7e-3) & (1.6e-3) & (0.47) & (0.43) & (0.24) & (0.19) & (2.2e-3) & (2.2e-3) & (0.59) & (0.56) & (0.40) & (0.31) \\
ALSTM+TRA& 0.107 & 0.102 & 60.27 & 59.09 & 57.66 & 55.16 & 0.119 & 0.112 & 60.45 & 59.52 & 59.16 & 58.24 \\
(multi-source) & (2.0e-3) & (1.8e-3) & (0.43) & (0.42) & (0.33) & (0.22) & (1.9e-3) & (1.7e-3) & (0.53) & (0.58) & (0.43) & (0.32) \\
\textbf{GRU-PFG} & \textbf{0.116} & \textbf{0.110} & \textbf{61.05} & \textbf{60.34} & \textbf{58.96} & \textbf{55.66} & \textbf{0.134} & \textbf{0.128} & \textbf{62.45} & \textbf{61.73} & \textbf{60.68} & \textbf{59.16} \\
(just Alpha360) & (2.5e-3) & (3.1e-3) & (0.74) & (0.66) & (0.48) & (0.30) & (3.2e-3) & (2.9e-3) & (1.08) & (0.69) & (0.39) & (0.35) \\
\hline
\label{table3}
\end{tabular}
\end{sidewaystable}

\begin{figure}[h]
\centering
\includegraphics[width=0.9\textwidth]{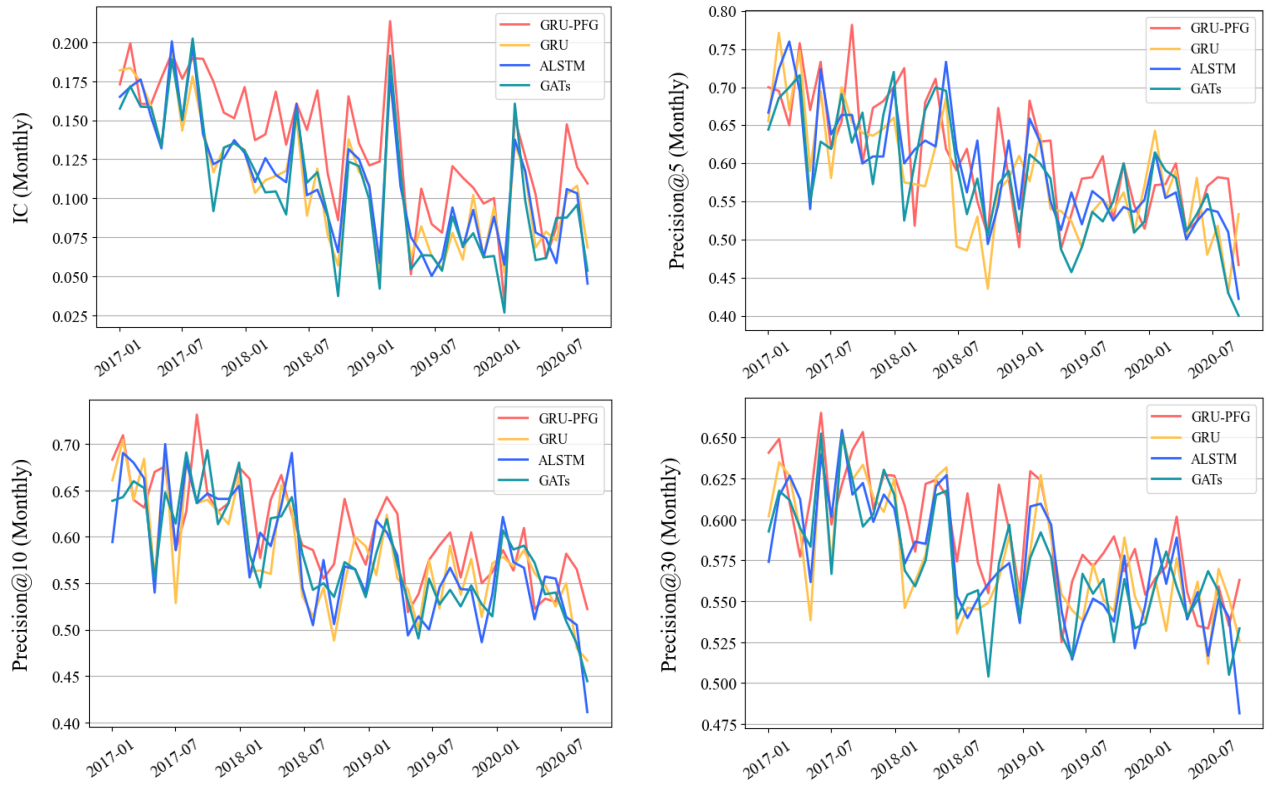}
\caption{Comparison of GRU-PFG with other models (just based on Alpha360) on monthly average IC and monthly average Precision@N at different time points, with metrics ordered from top left to bottom right as IC, Precision@5, Precision@10, and Precision@30}\label{fig3}
\end{figure}

Further, we compared the GRU-PFG model with several other models (just based on Alpha360) that performed well in prediction over time. Although models like ALSTM, GRU, and GATs have shown good performance on the CSI300 data, GRU-PFG model still maintains a significant advantage. Looking at the monthly average IC over different periods, GRU-PFG consistently leads by a large margin throughout most of the test period (See Figure \ref{fig3}), with only a few specific months where it is slightly surpassed by the other models.

In terms of the monthly average Precision@N metric, the advantage of GRU-PFG is not as pronounced as in IC. However, regardless of whether N is 5, 10, or 30, it is evident that the precision curves of GRU-PFG have higher upper and lower bounds, with the model maintaining a lead in most months.

\subsection{Impact of Model Parameters and Structure}

During the experimentation with our model, we made numerous macro and micro adjustments to the structure to achieve better prediction performance. To illustrate the adjustment process and the contribution of each structural component to the overall performance, we conducted ablation experiments to investigate the impact of different parts or sections.

From a macro perspective, our model consists of three main components: (1) Preliminary Information Extraction by GRU (2) Primary Relationship Extraction (3) Secondary Relationship Extraction. We compared the stock trend prediction performance with and without the latter two components. It was observed that as more components are added, the prediction performance improves progressively, especially after incorporating the second-level relationship extraction network, which significantly enhanced the prediction performance on the CSI300 dataset.

The effectiveness of our model in capturing stock relationships is a crucial foundation for its superior performance compared to previous models. On a micro level, capturing stock relationships relies on Pearson correlation in Equation \ref{equ4}. Although we also considered using cosine similarity to represent stock correlations, as referenced in other articles \cite{RN196, RN161}, it yielded relatively mediocre results. Both in terms of IC and Precision@30 metrics, the Pearson correlation coefficient demonstrated better prediction performance in terms of both upper limits and overall trends. This does not undermine the importance of cosine similarity in representing correlations, but for this specific task, Pearson correlation is more suitable for our model.

\begin{figure}[h]
\centering
\includegraphics[width=0.9\textwidth]{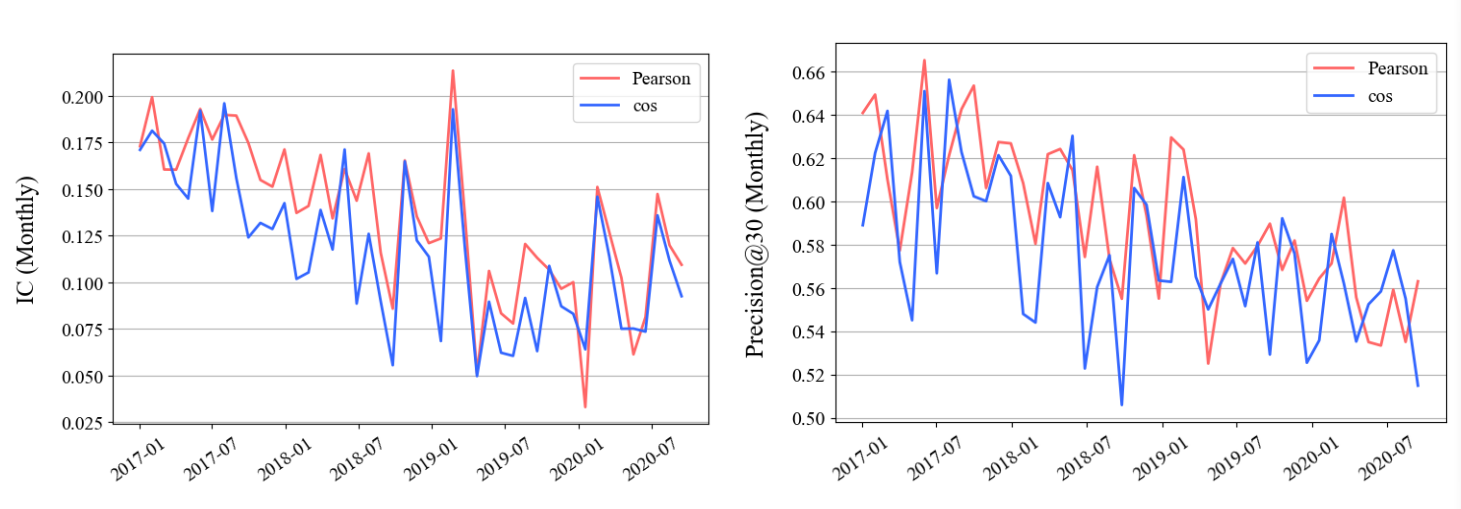}
\caption{Comparison of monthly average IC and monthly average Precision@N with different methods for measuring stock similarity}\label{fig4}
\end{figure}

\section{Conclusion}	
In this paper, we propose a new stock prediction model based purely on Alpha360 factors, called GRU-PFG. This model overcomes the limitations of previous models that were confined to factor relationships and achieved limited effectiveness. By representing the relationships between different stocks using a graph, it provides a new approach for stock prediction, resulting in a significant enhancement of evaluation metrics such as IC on the same dataset.

In the future, we hope to promote this model to help more investors achieve higher returns. Additionally, we aim to further improve the model to enhance its prediction performance across various datasets.

\bibliography{sn-bibliography}

\end{document}